# Characterization of atomization and delivery efficiency of exogenous surfactant in preterm infant lungs using an *ex vivo* respiratory model

Ghalia KAOUANE[1], Jean-François BERRET[2], Yannick CREMILLIEUX[3], Noël PINAUD[3], Fanny MUNSCH[4], Bei ZHANG[5], Michael FAYON[6], Rémy GÉRARD[6], Eric DUMAS DE LA ROQUE[7], Sophie PERINEL-RAGEY[8], Lara LECLERC[1], Jérémie POURCHEZ [1*]

[1] *Mines Saint-Etienne, Univ Lyon, Univ Jean Monnet, INSERM, U 1059 Sainbiose, Centre CIS, F - 42023 Saint-Etienne, France.*
[2] *Laboratory of Matter and Complex Systems, CNRS, UMR 7057, Université Paris Cité, Paris, France.*
[3] *Institute of Molecular Sciences, CNRS, UMR 5255, University of Bordeaux, France.*
[4] *Institute of Bioimaging, University of Bordeaux, Bordeaux, France.*
[5] *Canon Medical Systems Europe, Zoetermeer, The Netherlands.*
[6] *Bordeaux University Hospital (CHU Bordeaux), Department of Pediatrics, CIC-P INSERM 1401 & University of Bordeaux, France.*
[7] *Cardiothoracic Research Center of Bordeaux, INSERM U1045, F-33000 Bordeaux, France.*
[8] *Jean Monnet University, Saint-Étienne, INSERM SAINBIOSE U1059, F-42023 Saint-Étienne, France*

**ABSTRACT**
Administration of pulmonary surfactant is crucial for the treatment of respiratory distress syndrome (RDS) in preterm infants. The aim of this study is to evaluate the potential of Curosurf atomization *via* the Endosurf device, a recently developed spray technology, as a promising approach for surfactant delivery in infants with RDS. A comprehensive analysis was performed to evaluate the physicochemical properties of atomized Curosurf, including its surface tension and rheology. The size distribution of Curosurf vesicles was also investigated. An *ex vivo* respiratory model based on rabbit lungs breathing through an instrumented hypobaric chamber representing the thorax of a preterm infant was developed to provide proof of concept for regional aerosol deposition of atomized Curosurf. The atomization of Curosurf with the innovative Endosurf device did not significantly alter surface tension, but reduced vesicle size and promoted homogeneous distribution of Curosurf in the lungs. Rheological measurements showed the viscoelastic complexity of atomized Curosurf. This preliminary study confirmed the promising potential of Curosurf atomization *via* the Endosurf device for the distribution of surfactant in the lungs of infants with RDS. These advances could help to improve the treatment of RDS in preterm infants and offer new perspectives for healthcare professionals and affected families.

**\*CORRESPONDING AUTHOR**
Jérémie POURCHEZ
École Nationale Supérieure des Mines de Saint-Etienne
158 cours Fauriel, CS 62362
42023 Saint-Etienne Cedex 2 - FRANCE.
Email address: pourchez@emse.fr - Telephone number: +33477420180

# 1. Introduction

The alveoli are crucial for gas exchange in the lungs. They consist of around 480 million tiny vesicles in each lung, which form a huge surface area of 50 to 100 m² [1][2]. The alveoli are colonized by two cell types, type 1 pneumocytes (AT1) and type 2 pneumocytes (AT2), and are composed of lipids and surfactant-specific proteins [3][4]. Although they account for only 10 % of alveolar cells, AT1 cells cover 90 % of the alveolar surface and facilitate gas exchange by diffusion [5]. AT2 cells, which make up about 15 % of alveolar cells, produce pulmonary surfactant, which is mainly composed of phospholipids and proteins and forms a thin layer (100-500 nm thick) on the alveolar surface. This surfactant reduces surface tension, which is crucial for maintaining alveolar stability during respiration. The lipid component, which makes up about 90 % of the surfactant composition, consists mainly of phospholipids, with phosphatidylcholine being the most abundant, especially dipalmitoylphosphatidylcholine. Proteins make up about 10 % of the surfactant content. The surfactant synthesized by the AT2 pneumocytes first forms compact bilayers, the so-called lamellar bodies, which serve as a reservoir before it is released into the hypophase. The reduction of surface tension is favored by the interaction of lipids and proteins, which is crucial for the health of the lungs and the efficiency of gas exchange. Pulmonary surfactant can be inactivated by several mechanisms that contribute to or result from various diseases [6].

*In vivo*, the presence of pulmonary surfactant equalizes pressure distribution and prevents atrophy of smaller alveoli and hypertrophy of larger alveoli (Supplementary Data S1). This balance promotes effective gas exchange by mitigating the effects of capillarity and maintaining alveolar stability, which is critical for respiratory function. Premature births represent a major challenge for neonatology. They affect infants born between 22 and 37 weeks of gestation, as defined by the World Health Organization [7]. In the mid-twentieth century, the United States struggled with a high infant mortality rate due to preterm birth, which amounted to over 10,000 deaths per year. This led to the discovery of hyaline membrane disease, now known as acute respiratory distress syndrome (ARDS), by Avery and Mead in 1959 [8, 9]. ARDS manifests as respiratory distress causing arterial hypoxia [10]. Surfactant, which is crucial for efficient breathing, reduces the surface tension of the alveoli. Its production usually begins around the 24th week of gestation and reaches a functional level in the 34th to 36th week. However, premature infants often suffer from surfactant deficiency, which increases their risk of ARDS after birth [11]. The incidence of ARDS increases with prematurity. It affects around 1 % of all births, but increases to up to 90 % in children born before 28 weeks' gestation [12].

Endogenous pulmonary surfactant is obtained by bronchoalveolar lavage, which is performed in living mammals by bronchoscopy, an invasive procedure, or in sacrificed animals. Apart from ethical considerations, the quantities of endogenous surfactant obtained are small, and a damaged lung can lead to contamination with cellular material [6]. This complexity necessitates the use of exogenous pulmonary surfactant as an alternative, which is obtained from various sources. Commonly used animal surfactants include beractant "Survanta" and calfactant "Infasurf" from bovine lung and poractant alfa "Curosurf" from porcine lung. These substitutes differ not only in lipid and protein concentration, but also in vesicle structure [13,14,15,16] and rheology [17,18,19,20,21,22,23]. Curosurf has a higher concentration of phospholipids and requires a lower administration volume and is therefore preferentially used in hospitals for the treatment of ARDS [24].

Techniques for administering exogenous surfactant have evolved to better meet the needs of preterm infants with ARDS. Originally, surfactant was administered to mechanically ventilated preterm infants, but this often resulted in prolonged ventilation and did not effectively reduce bronchopulmonary dysplasia [25]. Currently, the most commonly used method is instillation of surfactant and rapid extubation [26]. Despite its popularity, instillation of surfactant can lead to problems during extubation, resulting in prolonged mechanical ventilation [27]. An alternative approach, known as less invasive surfactant administration (LISA), involves administering surfactant via a feeding tube or small catheter while the infant is breathing spontaneously. This method minimizes the need for invasive ventilation. Studies have shown that LISA can shorten the duration of mechanical ventilation and reduce the incidence of chronic lung disease in preterm infants [28]. Ongoing research aims to develop even less invasive methods such as nebulization or surfactant nebulization. Nebulization offers potential benefits, including more uniform distribution and less reliance on mechanical ventilation. However, further clinical studies are needed to fully evaluate these techniques [25]. Preliminary results from a pilot study suggest that delivery of aerosolized surfactant via nasal continuous positive airway pressure (nCPAP) is a viable and safe option [29].

The aim of this work is to characterize and evaluate the performance of atomization of exogenous surfactant using an innovative spray technology to overcome the challenges in the treatment of preterm infants with ARDS. A team of neonatologists, pediatric pulmonologists and French start-up companies have developed the patented concept of the Endosurf medical device [30]. This innovative device aims to (i) avoid the painful laryngoscopy required for the LISA technique, (ii) deliver surfactant by atomization to achieve a better distribution in the lung, and (iii) provide a more user-friendly learning curve compared to the LISA technique. In this study, we will investigate whether atomization with the Endosurf technology changes the physicochemical properties of the exogenous surfactant such as surface tension, rheology, vesicle size and structural properties of the Curosurf vesicles. In addition, proof of concept for regional aerosol deposition of surfactant in the lung is provided using an *ex vivo* respiratory model based on a heart-connected rabbit thorax breathing through an instrumented hypobaric chamber mimicking a preterm infant.

## 2. Materials and Methods
**Endosurf medical device**
Atomized surfactant is intended for endotracheal administration in the treatment of neonatal respiratory distress syndrome. In our study, we used the patented concept of the Endosurf medical device **[30]** to atomize the surfactant Curosurf (Chiesi Farmaceutici, Parma, Italy). The principle of the Endosurf device is to introduce the surfactant under pressure through a nozzle to produce a spray aerosol. The nozzle has a complex, previously analyzed geometry that produces a fine spray while avoiding clogging and maintaining sufficiently low pressures (below 100 bar) to avoid mechanical damage to the nozzle. The entire Endosurf device, including the syringe pump containing the surfactant, is kept at 37 °C both before and during spraying. This temperature control ensures that the surfactant remains under physiological conditions throughout the process, maintains its physico-chemical properties and accurately replicates the clinical application. To exert controlled pressure during atomization, the Endosurf was coupled with a 3340 series universal testing system (INSTRON, Élancourt, France), which enables precise linear movement at a constant speed of 500 mm/min over a distance of 120 mm.

**Physicochemical features of exogenous surfactant pre- and post-atomization**

Before atomization, the Curosurf was heated to 37°C to reach the physiological temperature. Three different devices were used for the tests, each operating at a specific atomization pressure. Each Curosurf vial, lot number 1154081, initially contained 2.5 mL of product, with 0.5 mL reserved as a blank control (**Table 1**). After atomization, the samples, whether nebulized or not, were stored at 4°C for subsequent analysis of their physicochemical properties.

*Table 1: Summary Table of Atomized / Non-atomized Curosurf Volumes and pressure.*

| Endosurf device number | Pressure applied (bar) | Batch number of Curosurf | Surfactant volume (mL) | Volume of surfactant loaded in the syringe (mL) | Volume of recovered surfactant after atomization (mL) | Non-atomized Curosurf (mL) |
|---|---|---|---|---|---|---|
| (1) | 30.9 | 1154081 | 3 | 2.5 | 2 | 0.5 |
| (2) | 59.5 | | 3 | 2.5 | 2 | 0.5 |
| (3) | 75.8 | | 3 | 2.5 | 2 | 0.5 |

There are different approaches to evaluate the change in physicochemical properties of Curosurf intended for administration to preterm infants: measurement of surface tension, rheological analysis, structure and size of surfactant vesicles.

**Surface tension measurement**

One of the fundamental physicochemical properties of surfactants is their ability to regulate the surface tension between alveoli and air. In this study, the surface tension of exogenous surfactant solutions (Curosurf) was evaluated in both atomized and non-atomized forms using the hanging drop method. Measurements were performed at room temperature (+20°C) using the Digidrop DX device (GBX Scientific LTD, Romans-sur-Isère, France), which allows the evaluation of contact angle, wettability, surface energy and surface tension. Surface tension was measured for each atomized Curosurf sample (Endosurf devices (1), (2) and (3)) in three replicates and for pooled, non-atomized Curosurf in six replicates (**Figure 1**). The droplet profiles, including volume, angle and base diameter, were analyzed. Surface tension values were calculated based on the known density of Curosurf at +20°C (1.0018 g/cm³) and the collected data. After measurement, the Curosurf droplets were collected in a vial and stored at +4°C for subsequent rheological analysis.

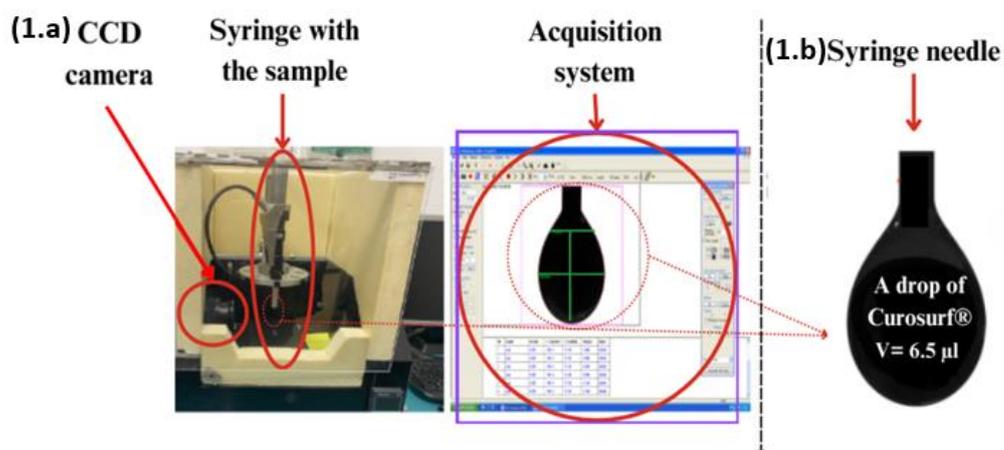

*Figure 1: Measurement of surface tension using the hanging drop method. (1. a) light source, CDD camera, a platform and a dosing unit with a syringe. (1. b) Image of a Curosurf pendant drop in air attached to the syringe needle.*

**Cone-and-Plate rheology**

The rheological experiments were carried out using a RheoCompass MCR 302 rheometer (Anton Paar SAS, Courtaboeuf, France) equipped with a cone and a plate with a diameter of 50mm and a cone angle of 1°, as well as a solvent trap to minimize water evaporation, and operated using RheoCompass software (version 1.26, Anton Paar; available at https://www.anton-paar.com). Measurements of the complex modulus $G^*(\omega) = G'(\omega) + iG''(\omega)$, where $G'(\omega)$ and $G''(\omega)$ are the elastic and loss moduli, were recorded as a function of the strain $\gamma_0$ and the angular frequency $\omega$ through strain and frequency sweeps, respectively. In strain sweeps, the angular frequency was held constant at 1 rad s$^{-1}$ while varying the applied strain $\gamma_0$ from 0.1 to 100%. For all Curosurf samples tested, we found a linear regime below the deformation $\gamma_0$ = 10%. Moduli were hence measured at the slightly lower strain of 5%. To evaluate measurement reproducibility over time, strain and frequency sweep tests were first performed on freshly opened Curosurf batches. Once loaded into the cone-and-plate geometry, each sample was measured sequentially two to three times. In frequency sweeps, the angular frequency was varied between 0.1 and 100 rad s$^{-1}$. Both atomized and non-atomized Curosurf samples were evaluated following the same experimental procedures. Each sample's cone-and-plate measurements were conducted in duplicate at a temperature of +37°C (**Figure 2**).

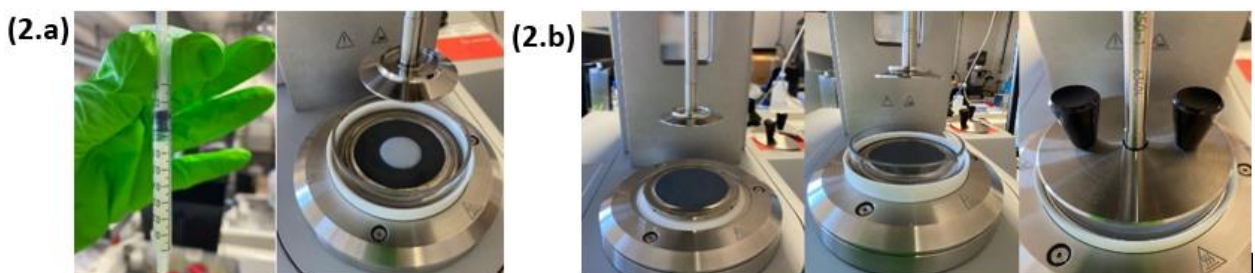

*Figure 2: Measuring the viscosity of atomized and not atomized Curosurf.*
*(2. a) cone and plate geometry with a diameter of 49.957mm and a cone angle of 0.997, a sample volume of 0.65mL was used. (2.b) Accessories were used to prevent evaporation of the samples, which were stored at a temperature of +37°C.*

**Optical microscopy**

For direct visualization of the Curosurf vesicles, an inverted microscope IX73 (Olympus SAS, Rungis, France) with an x60 objective (numerical aperture 0.70) was used, allowing brightfield and phase-contrast imaging [31, 32]. The data acquisition system consisted of an EXi Blue CCD camera (QImaging, Surrey, Canada) with Metamorph (version 7.10.3, Universal Imaging Inc., Bedford Hills, USA; available at https://www.moleculardevices.com/products/cellular-imaging/metamorph-research-imaging). For this study, four samples of the exogenous surfactant were analyzed, including one sample before atomization and three samples after atomization at different magnifications (x20, x40 and x60). In the method, 2 µl of the sample was placed between two 16 mm diameter round coverslips (**Figure 3.a, 3.b**) held together with a homemade screw clamp system (**Figure 3.c**). A stream of air, which was directed into the measuring cell through an air inlet cover, was used to heat the sample to +37°C. The vesicle images were digitized and processed using ImageJ Fiji software (version 2.9.0, 2022-09-14, available at https://fiji.sc/) and plugins **[33]** for structural analysis, including vesicle size determination. Vesicle quantification was performed by counting the number of vesicles within image areas of identical size, randomly selected across all samples. For size distribution analysis, the diameters of 250 vesicles per condition were manually measured from representative image fields. The resulting data were analyzed and graphically represented using GraphPad Prism software (version 9.5.0, GraphPad Software, Boston, USA; available at https://www.graphpad.com/). For size analysis, only phase-contrast images acquired at ×60 magnification were used to ensure optimal visualization and accurate individualization of the vesicles.

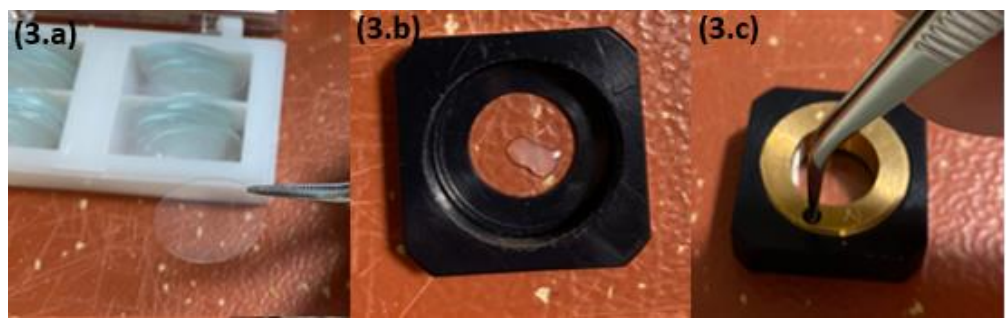

***Figure 3:*** *Microscopic Observation and Analysis of Samples Using Phase-Contrast (PC) and Bright-Field (BF) Techniques. (3.a) 16 mm diameter round coverslip employed as the sample substrate; (3.b) 2 µL of Curosurf carefully deposited onto the coverslip for analysis; (3.c) screw clamp system used to securely mount the sample during observation.*

**Aerodynamic parameters of atomized surfactant using the SPRAYTEC device**

We measured the particle size distribution (PSD) of aerosol droplets using the SPRAYTEC laser diffraction technique (Malvern Panalytical, Palaiseau, France; software version 4.0, available at https://www.malvernpanalytical.com). An open laser beam (range 0.5-900µm) generated with a 300F lens –was used to analyze the droplets after 15 seconds of atomization. The laser diffraction instrument determined the size distribution by measuring the angular variation of the scattered light intensity, creating a diffraction pattern. The PSD was characterized by the mean mass diameter (D (0.5)), which indicates the size at which 50% of the sample is smaller and 50% is larger. The values given are averages of at least three determinations. Nine Endosurf devices were used to determine the PSD of the generated spray (**Table 2**), using the same method as in the previous section (Physicochemical properties of the exogenous surfactant before and after atomization).

***Table 2:*** *Summary Table of Atomized Curosurf Volumes and Pressure.*

| Endosurf device number | Pressure applied (bar) | Batch number of Curosurf | Surfactant volume (mL) | Volume of surfactant loaded in the syringe (mL) | Volume of recovered surfactant after atomization (mL) | Non-atomized Curosurf (mL) |
|---|---|---|---|---|---|---|
| (4) | NA | 1170710 | 3 | 2.5 | 2 | 0.5 |
| (5) | 20 | 1154081 | 3 | 2.5 | 2 | 0.5 |
| (6) | 27.5 | 1154081 | 3 | 2.5 | 2 | 0.5 |
| (7) | 30 | 1170710 | 3 | 2.5 | 2 | 0.5 |
| (8) | 35 | 1170710 | 3 | 2.5 | 2 | 0.5 |
| (9) | 35 | 1163893 | 3 | 2.5 | 2 | 0.5 |
| (10) | 35 | 1162353 | 3 | 2.5 | 2 | 0.5 |
| (11) | 35 | 1162353 | 3 | 2.5 | 2 | 0.5 |
| (12) | 40 | 1162353 | 3 | 2.5 | 2 | 0.5 |

## *Ex vivo* respiratory model and MRI imaging of surfactant distribution in the lung

The identification of deposition sites plays a crucial role in the assessment of tissue doses and subsequent biological effects. An innovative methodological approach based on *ex vivo* imaging experiments was chosen. Anatomical models of animals intended for human consumption were used (**Figure 4.a**). These experiments were carried out on anatomical rabbit models obtained from food industry waste. They were Hyplus rabbits, a commercial hybrid line commonly used in the meat industry (Hypharm, France **[34]**), slaughtered at a weight of 3 kg at an age of 84 to 90 days. The dissected thoraxes were placed in a hypobaric chamber equipped with instruments mimicking pleural depression (**Figure 4.b**) **[35]**. This *ex vivo* model is fully consistent with the 3R principles — Replacement, Reduction and Refinement — a 50-year-old ethical framework that aims to minimize animal suffering, increase the reliability of scientific data and promote the use of replacement methods for live animal experiments. By relying exclusively on biological material from the food industry rather than using live animals, our approach is in line with these guidelines and offers a relevant, ethical and scientifically sound alternative to conventional animal experimentation. Thus, the lungs were ventilated *in vivo* similar to passive ventilation. In addition, a neonatal ventilator (Babylog, Draeger, Lübeck, Germany) was connected to the trachea to maintain a continuous positive airway pressure of 6 cmH2O. A solution of exogenous surfactant and contrast agent was prepared by mixing Curosurf and a gadolinium-based contrast agent (Dotarem, Guerbet, Villepinte, France) in a 10:1 volume ratio (2.7 ml Curosurf and 0.27 ml contrast agent). The solution was heated in a water bath at +37°C for 5 to 10 minutes. Subsequently, 2.5 ml of the solution was placed in the Endosurf device at body temperature. The device was inserted into the trachea of the lung and the solution was nebulized within a few seconds **(Figure 4.c)**. The chamber was then placed in the center of the MRI magnet for the acquisition of MRI images (**Figure 4.d**). MRI scans were also performed on an isolated rabbit thorax after 2.7 ml of surfactant solution and 0.27 ml of gadolinium-based contrast agent were mixed in a 10:1 volume ratio, pre-warmed to +37°C, and instilled into the rabbit lung using the LISA reference method. MRI scans were performed using a clinical 3T whole-body magnet (Vantage Galan 3T ZGO, Canon Medical Systems Corporation, Japan). MR images were acquired using a 3D MRI sequence with ultra-short reverberation time (UTE) and the following acquisition parameters: Repetition time = 3.7ms, echo time = 96µs, 1 mean, total acquisition time = 3min28sec, field of view = 11.3x11.3cm2, slice thickness = 1mm, voxel size = 0.78x0.78x1mm3 **[36]**.

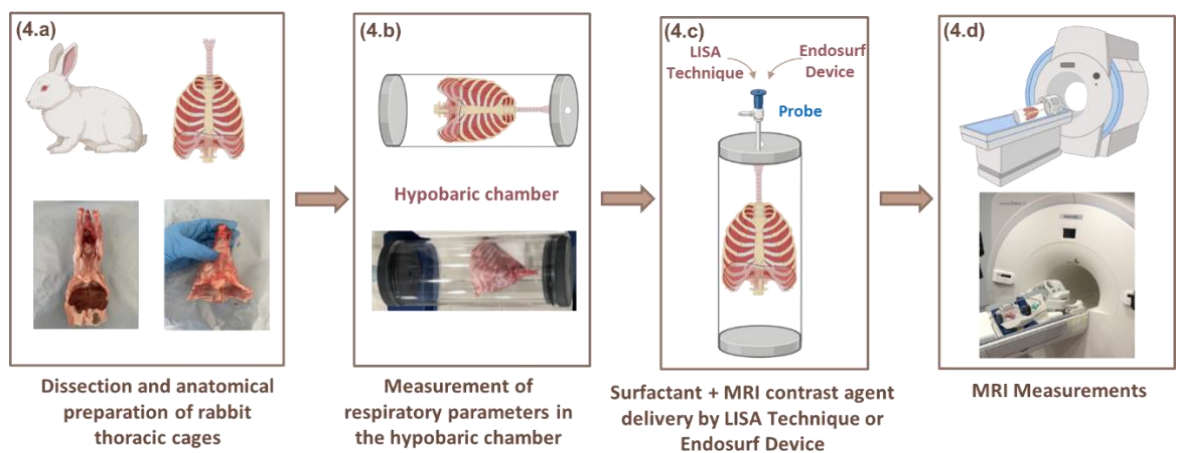

***Figure 4:*** *Ex vivo respiratory model in a hypobaric chamber and MRI imaging of surfactant distribution in rabbit lungs.*

## 2.1. Statistical analysis

Statistical analyses were performed with GraphPad Prism (version 9.5.0). Data are presented as mean ± standard deviation. For comparisons between groups, either the unpaired Student's t-test or the Mann–Whitney U-test was used, depending on the data distribution. A p-value < 0.05 was considered statistically significant. The number of replicates for each experiment is indicated in the corresponding figure legends or method descriptions.

# Results and Discussion

Comparison of physicochemical features of exogenous surfactant pre- and post-atomization with the Endosurf spray device

**Surface tension**

The results show an average surface tension of 30.0 ± 1.405 mN m-1 for the non-atomized surfactant and 31.3 ± 1.46 mN m-1 for the atomized surfactant (**Figure 5**). An unpaired Student's t-test revealed no statistically significant difference between the two groups (p = 0.3156), indicating that atomization did not significantly alter the surface properties of the exogenous surfactant. In addition, the coefficient of determination ($R^2$ = 0.07732) shows that only a small part of the variance in surface tension is explained by atomization. This weak correlation confirms that atomization has only a minimal effect on the surface tension of the surfactant. However, it should be noted that despite a non-significant difference between non-atomized and atomized surfactants, the standard deviation in the atomized group remains high. When analyzing the results, we could not identify any particular behavior that would explain this observation as a function of the pressures measured for each Endosurf medical device (e.g., it is not the devices with the highest pressures that produce the lowest or highest surface tension values).

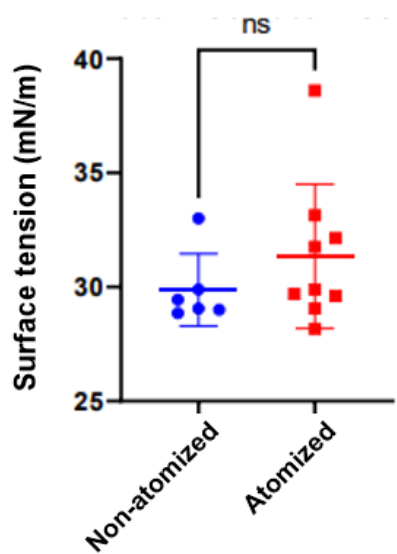

***Figure 5:*** *Comparison of the surface tension (mN/m) between atomized and non-atomized exogenous surfactant. Surface tension of each drop is plotted with the error bars representing the standard deviations.*

**Rheological properties**

Strain and frequency sweep measurements were first performed on freshly opened Curosurf batches. Once loaded into the tool geometry, the samples were measured sequentially two or three times to test the reproducibility of the measurements over time. **Figure 6.a** shows the frequency dependence of the elastic and loss moduli for two freshly opened samples (batch number 1162353). Under the used conditions, the measurement was made within minutes after opening the flask. In **Figure 6.a** it can be seen that the two measurements gave similar results, with data points largely superimposed over the entire frequency range. A second observation is that the elastic and loss moduli are close to each other, with values of the order of 0.05Pa at 1rad $s^{-1}$. This indicates viscoelastic behavior, with a low elastic contribution. The rheological response shows power-law dependences of the form $G'(\omega) \sim \omega^\alpha$ and $G''(\omega) \sim \omega^\beta$, with exponents $\alpha$ = 0.38 ± 0.02 and $\beta$ = 0.56 ± 0.02. In line with the results obtained for fluids undergoing a sol-gel transition, the exponent values indicate that Curosurf at 80 g $L^{-1}$ is below the sol-gel transition, which is characterized by a rheological state where $G'(\omega) > G''(\omega)$ with both $\alpha$ and $\beta$ of the order of 0.7 **[37-39]**. Compared with the data from Ciutara et al. who also carried out elastic and loss modulus measurements on Curosurf **[40]**, $G'(\omega)$ and $G''(\omega)$ values measured here are lower by a factor of 2. It should be added that Ciutara *et al.* **[40]** also found $G'(\omega) > G''(\omega)$ over the 0.5-10rad $s^{-1}$ measurement range, leading the authors to conclude that Curosurf at 80g.$L^{-1}$ was in a gel state.

Regarding the linear rheology of Curosurf, our results differ slightly from those reported by Ciutara et al. **[40]** and Thai et al. **[32,41]**. Specifically, we observe a crossover of the G'(ω) and G''(ω) curves around 3 rad $s^{-1}$, with G'(ω) > G''(ω) at lower frequencies and G''(ω) > G'(ω) at higher frequencies. Taken together, these results indicate some variability in the viscoelastic moduli between Curosurf batches, which could be attributed to differences in the product synthesis. To determine the long-term stability of Curosurf physical properties over time, and in particular their viscosity properties, $G'(\omega)$ and $G''(\omega)$ were measured on the same sample shortly after opening the bottle, after one day, and after one month (**Figure 6.b**). Between these measurements, the sample was kept at +4°C in the dark. We observed that after one day the results were similar to those for the freshly opened sample. However, after one month, there was a significant change in the frequency behavior of the complex elastic modulus $G^*(\omega)$, with the elastic modulus increasing by a factor of 20 and the loss modulus by a factor of 10 with respect to the initial sample. Additionally, $G'(\omega)$ exceeded $G''(\omega)$, and the frequency behaviors had exponents $\alpha$ = 0.07 and $\beta \sim 0$, clearly indicating a gel-like behavior **[37,42]**. These measurements suggest that the aging effects of Curosurf occurred over a time scale of one month, and that precautions should be taken when measuring exogenous pulmonary surfactants **[32,37,40,41,43-46]**.

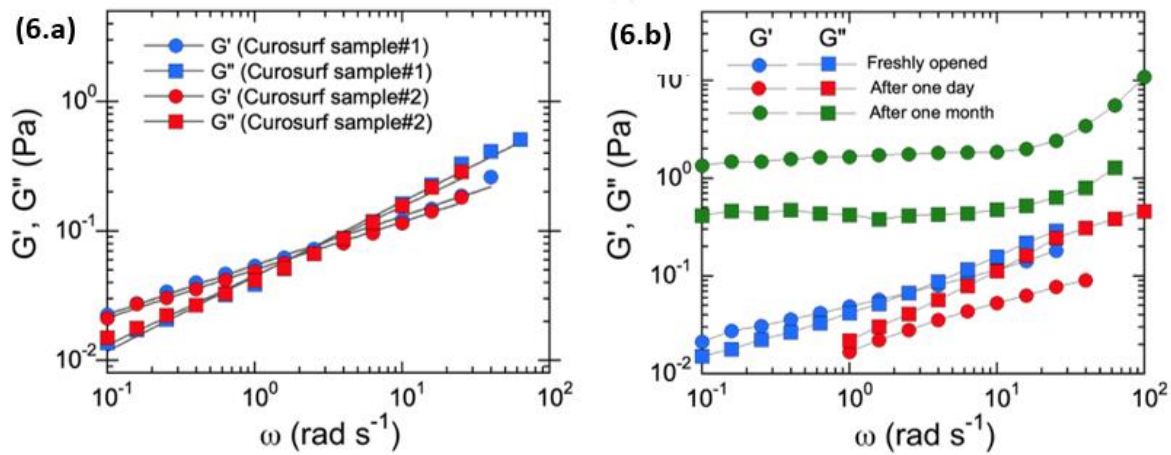

*Figure 6: Elastic and loss moduli of freshly opened Curosurf samples as a function of the angular frequency (T = +37°C). The cone-and-plate measurements were performed in the linear regime of deformation ($\gamma_0$ = 5%). The straight lines through the data points result from best-fit calculations using power-law functions of the form $G'(\omega) \sim \omega^\alpha$ and $G''(\omega) \sim \omega^\beta$, with exponents α = 0.38 and β = 0.56. B) $G'(\omega)$ and $G''(\omega)$ obtained from a Curosurf sample shortly opening, after one day and after one month. After opening, the dispersion was kept at +4°C in the dark. The plateau in modulus $G'(\omega)$ after one month suggests viscoelastic gel behavior.*

**Figure 7** compares the linear viscoelastic behaviors of samples originating from the same batch (number 1154081) atomized using different versions of the Endosurf prototypes. These prototypes are termed Device (1) (**Figure 7.a-b**), Device (2) (**Figure 7.c-d**), and Device (3) (**Figure 7.e-f**) respectively. The upper panels show the elastic moduli, while the lower panels display the loss modulus. For most data collected, $G'(\omega)$ and $G''(\omega)$ exhibit moduli around 0.05–0.1Pa, similar to those of freshly opened samples. The continuous blue lines for $G'(\omega)$ and red lines for $G''(\omega)$ represent the power laws obtained from freshly opened flasks (**Figure 6.a**), with α and β being 0.38 and 0.54, respectively. The data obtained before and after atomization show good agreement, except for those in **Figure 7.c** at high frequencies. In some cases, there is even an overlap between non-atomized and atomized samples (**Figure 7.b, 7.e, 7.f**). These data indicate that the rheological properties of the atomized Curosurf samples are similar to those of the initial samples, confirming the potential use of Endosurf devices for the delivery of exogenous surfactants to humans.

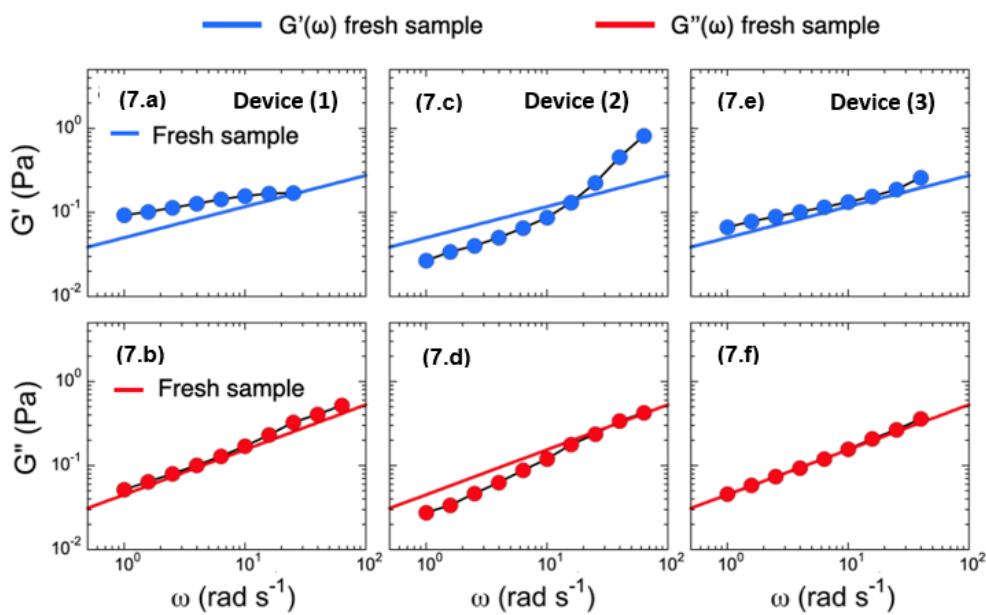

***Figure 7**: Elastic (upper panels) and loss (lower panels) modulus of a Curosurf sample atomized through 3 different Endosurf prototypes. These prototypes are labelled Device (1) (a-b), Device (2) (c-d), and Device (3) (e-f) respectively. The experiments were carried out in the linear regime of deformation ($\gamma_0$ = 5%) and at +37°C. The solid lines show the results for the non-atomized sample, for comparison.*

## Structure and size of surfactant vesicles
### Phase contrast microscopy

Phase contrast microscopy was used to evaluate the effects of atomization on the morphology of the exogenous surfactant vesicles (Supplementary Data **S2**). The main objective was to compare the morphological features and size profiles of the vesicles between the non-atomized exogenous surfactant (**Figure 8 - image a**) and the atomized counterparts (**Figure 8 - images b, c and d**).

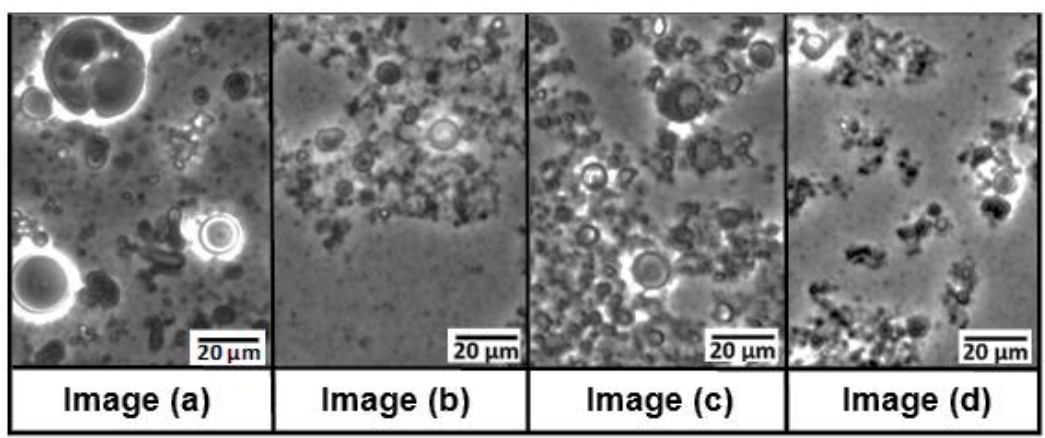

***Figure 8**: Microscopic images of exogenous surfactant. Fresh, non-atomized exogenous surfactant (image a) and atomized with Endosurf devices (1), (2) and (3) (images b, c and d). The images were taken with the Olympus IX73 microscope with phase contrast and processed with ImageJ Fiji software (magnification ×60, scale bar = 20 μm).*

Visual inspection of the microscopic images reveals a difference in the size of the vesicles between the non-atomized exogenous surfactants and their atomized counterparts. To quantitatively delineate this difference, precise measurements of the average vesicle size were performed in representative image areas (**Table 3**). The results show that the average size of the vesicles in the non-atomized exogenous surfactants is about 2.2 µm, which is almost twice as large as that of the atomized vesicles, which is about 1.1 µm.

*Table 3*: *Summary of measurement of diameter/number of exogenous surfactant vesicles.*

| Images | Number of vesicles | Mean ± SD (µm) | Mode ± SD (µm) |
|---|---|---|---|
| (a) Non-atomized | 273 | 2.2 ± 2.5 | 0.65 ± 2.5 |
| (b) Device 1 | 781 | 1.0 ± 0.8 | 0.62 ± 0.8 |
| (c) Device 2 | 327 | 1.3 ± 1.3 | 0.44 ± 1.3 |
| (d) Device 3 | 685 | 1.0 ± 0.9 | 0.50 ± 0.9 |

Other remarkable observations concern the frequency of the vesicles. The number of small vesicles is significantly higher in the atomized samples than in the non-atomized samples. If we compare Image (a) and Image. (b), we can see that although the surface area covered by the vesicles remains more or less the same, the number of vesicles increases. Image (a) shows about 273 vesicles, while Image (b) shows 781 vesicles (about 2.8 times more). This increase in vesicle number can be explained by the fragmentation of larger vesicles into smaller ones induced by the atomization process. The mechanical forces and high pressure applied during atomization likely break larger vesicles into multiple smaller ones while conserving the total surfactant volume. This phenomenon is consistent with volume conservation principles widely reported in aerosol and emulsion science **[47, 48].**

**Analysis of exogenous surfactant pre- and post-atomization**

The results show an average vesicle size of 2.2 ± 2.5 µm for non-atomized Curosurf and 1.080 ± 0.9649 µm for atomized Curosurf (**Figure 9**). A Mann-Whitney U test confirmed a statistically significant reduction in vesicle size after atomization ($P < 0.0001$), indicating that the process effectively breaks up larger vesicles into smaller structures.

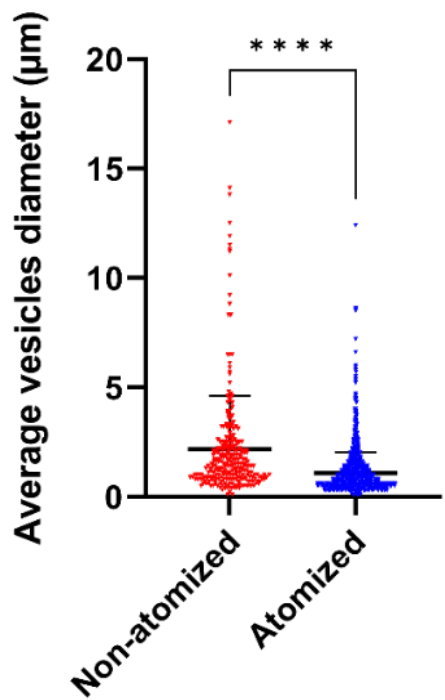

*Figure 9:* Average vesicles diameter of exogenous surfactant before and after atomization

In addition to the reduction in mean size, the median vesicle size also decreased from 1.4 µm to 0.8 µm, with an absolute difference of 0.6 µm, demonstrating the significant effect of atomization on vesicle size. The maximum vesicle size decreased from 17.1 µm to 12.4 µm, and the overall size range shrank from 17.1 µm to 12.4 µm, suggesting that atomization not only reduces vesicle size, but also limits the presence of larger vesicles that might otherwise interfere with effective deposition in the lung. A notable reduction in size variability was also observed, as evidenced by a decrease in standard deviation from 2.5 µm (non-atomized) to 1 µm (atomized). This greater uniformity in vesicle size could contribute to improved aerosol stability and a more uniform distribution of surfactant in the pulmonary system. These results confirm that atomization significantly reduces vesicle size, improves homogeneity and minimizes variability - factors that are critical for optimizing aerosol performance, surfactant distribution and deposition efficiency. These structural changes may ultimately improve the clinical efficacy of surfactant replacement therapy by ensuring better coverage and distribution in the lungs.

**Distribution of vesicle sizes of exogenous surfactants pre- and post-atomization**
This study presents the results of analyzing the size distribution of exogenous surfactant vesicles, where a total of 250 vesicles were examined for each sample condition (see Supplementary Data **S3**). The aim was to compare the size distribution between non-atomized and atomized samples. **Figure 10** shows significant differences in the size distribution of the vesicles before and after atomization. The mean vesicle diameter for the non-atomized sample is 1.33 µm, accompanied by a wide distribution range extending up to 6 µm, indicating considerable heterogeneity within the non-atomized sample. In comparison, the mean diameters of the atomized samples were significantly smaller, with values between 0.67 µm and 0.84 µm for the three different atomization devices (devices 1, 2 and 3).

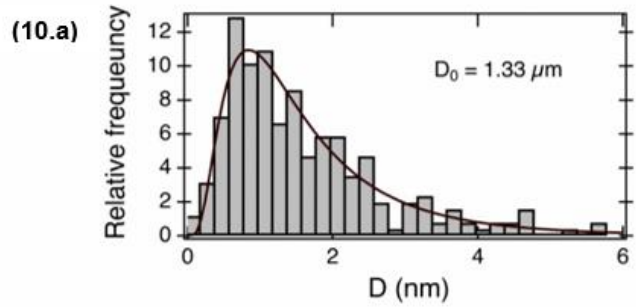

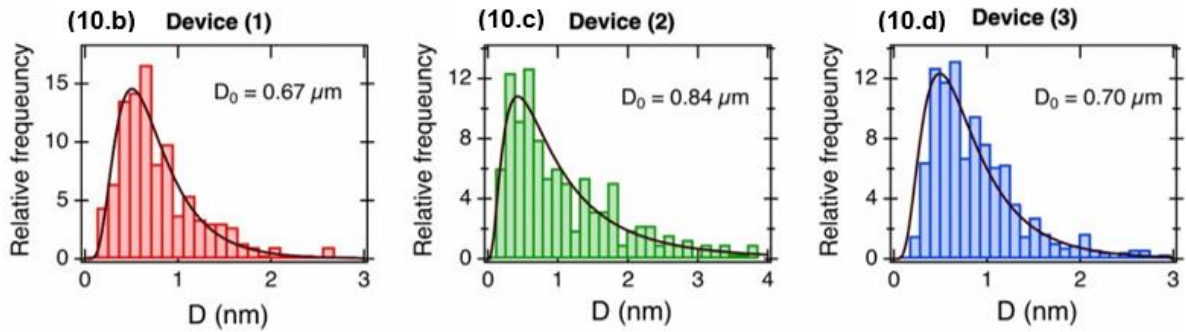

*Figure 10*: Distribution of the vesicle size of the exogenous surfactant before and after atomization. *(10.a)* Relative frequency distribution of vesicle diameters (D) in the non-atomized sample, with a mean diameter $D_0 = 1.33 \mu m$. *(b–d)* Relative frequency distributions of vesicle diameters for three different atomization devices: *(10.b)* Device (1), $D_0 = 0.67 \mu m$; *(10.c)* Device (2), $D_0 = 0.84 \mu m$; and *(10.d)* Device (3), $D_0 = 0.70 \mu m$.

The atomized samples consistently had smaller vesicle diameters than the non-atomized samples. For example, the average vesicle diameter of the sample atomized with Device 1 was 0.67 µm, highlighting the effectiveness of atomization in reducing vesicle size. In addition, the atomized samples exhibited lower standard deviations, indicating a more uniform size distribution of the measured vesicles. Optical phase contrast microscopy also showed remarkable differences in the morphology and size distribution of the vesicles. The non-atomized vesicles had an average size of 1.33 µm, while the average size of the atomized vesicles was significantly smaller at about 0.75 µm. This drastic reduction in size illustrates the mechanical fragmentation that occurs during atomization, whereby larger vesicles are broken down into smaller particles. Overall, the results indicate that atomization produces smaller and more uniform vesicles, which are beneficial for aerosol delivery. Our study found that the non-atomized vesicles were on average 1.3 µm in size, which differs slightly from the 3.3 µm reported in the literature **[32]**. This discrepancy could be due to the use of phase contrast microscopy, which can more accurately detect smaller vesicle diameters. Despite this difference, our results are consistent with existing knowledge on non-atomized surfactant vesicles, confirming the reliability of our results. This reduction in size and improvement in homogeneity may potentially increase the clinical efficacy of surfactant therapies, particularly in the treatment of respiratory diseases. Our analysis is consistent with the existing literature on non-atomized surfactant vesicles and confirms the reliability and relevance of our findings.

**Aerodynamic parameters of atomized surfactant**

We monitored the PSD and transmittance of the exogenous surfactant aerosol generated by different Endosurf devices in real time. Initially, high transmittance values indicated low particle generation. Then atomization began and transmittance decreased significantly, indicating aerosol generation. Finally, when atomization was complete, the transmittance values stabilized and then increased, indicating insufficient particle production. The transmission curve versus time showed a U-shape throughout the process. For each measurement, the "Selected Zone" option was used to focus on the relevant data and exclude the values at the beginning and end of atomization. An example of this behavior is presented in supplementary data **S4**, where the red curve illustrates the typical transmittance profile, while the colored curves show the evolution of droplet diameters ($D_{10}$, $D_{50}$, $D_{90}$) over the same period. For each measurement, the "Selected Zone" option was applied to isolate the central portion of the atomization event and exclude transitional values at the beginning and end.

**_Table 4:_** *Results of $D_{10}$, $D_{50}$ and of $D_{90}$ after Spraytec analysis. $D_{10}$ (µm), $D_{50}$ (µm), and $D_{90}$ (µm) were calculated using the Spraytec software. They represent maximal particle size diameter that includes 10%, 50% and 90% of total particles volume, respectively.*

| Device | Caption | $D_{10}$ (µm) | $D_{50}$ (µm) | $D_{90}$ (µm) |
|---|---|---|---|---|
| **(4)** | | 74.44 | 146.53 | 352.68 |
| **(5)** | | 122.29 | 226.47 | 441.29 |
| **(6)** | | 99.30 | 180.75 | 394.27 |
| **(7)** | | 97.05 | 193.96 | 432.27 |
| **(8)** | | 19.31 | 36.91 | 66.26 |
| **(9)** | | 22.49 | 43.45 | 80.05 |
| **(10)** | | 27.13 | 52.42 | 99.65 |
| **(11)** | | 25.50 | 56.46 | 120.58 |
| **(12)** | | 20.00 | 44.21 | 86.85 |

As shown in **Table 4**, the average $D_{10}$ values for these devices range from 19.31 to 122.29µm, the $D_{50}$ values range from 36.91 to 226.47µm and the $D_{90}$, values range from 66.26 to 432.27µm. $D_{50}$ and $D_{90}$, which reflect most of the aerosol particles produced by Endosurf devices, are the most commonly used parameters for evaluating aerosol particle size distribution.

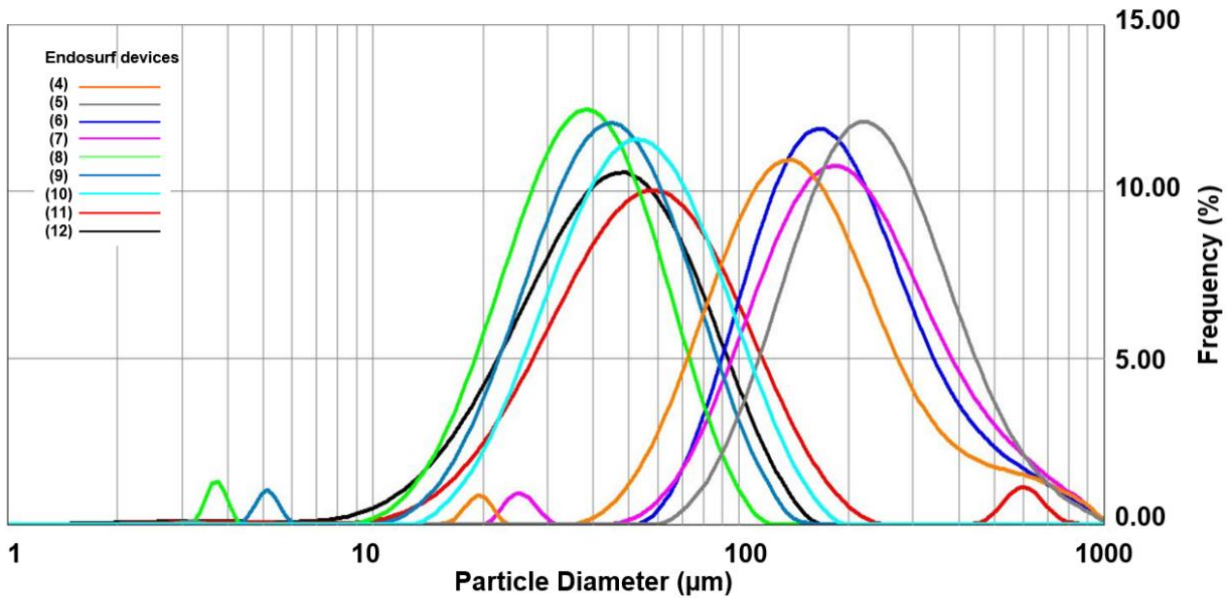

*Figure 11*: *Particle size distribution of aerosol droplets generated by different devices, measured with the Spraytec. The Spraytec system was used to evaluate the performance of nine different Endosurf devices. The key parameters, including $D_{10}$, $D_{50}$, $D_{90}$ and transmission, were automatically calculated using the Spraytec software. $D_{10}$, $D_{50}$ and $D_{90}$ represent the particle size diameters at which 10%, 50% and 90% of the aerosol particles are smaller, respectively. The particle size distribution (PSD) results for each device (labelled 4 to 12).*

The figure shows the particle size distribution of the aerosol droplets produced by nine different devices. Each curve in the diagram represents a different device, which is identified by a specific caption. The measurements show that the particle size distribution (PSD) results for each device (labeled 4 to 12) show significant differences, illustrating the differences in aerosol generation between the devices with two different droplet size distributions (**Figure 11**). Five devices (8, 9, 10, 11 and 12) produce droplets with an average diameter of 46.69 ± 6.9 µm, while the remaining four devices (4, 5, 6 and 7) produce droplets with a smaller average diameter of 186.93 ± 23.29 µm. These differences can be attributed to variations in the manual assembly of the nozzles. It is important to note that manual tightening, even when performed by the same person, can affect the size of the spray orifice. Tightening more can slightly reduce the diameter of the nozzle, resulting in the production of smaller droplets. We also observe a correlation between the mean droplet diameter and the applied pressure. In particular, a higher pressure leads to smaller droplets (*e.g.*, device 12: 40 bar pressures, mean diameter of 44.21 µm), while a lower pressure leads to larger droplets (*e.g.*, device 5: 20 bar pressure, mean diameter of 226.47 µm). To further evaluate the variability in particle size distribution (PSD) between the Endosurf devices, we categorized the $D_{50}$ values obtained from the SPRAYTEC measurements into two different groups based on their average droplet diameters: a high $D_{50}$ group (devices 4, 5, 6 and 7) and a low $D_{50}$ group (devices 8, 9, 10, 11 and 12). Each group comprised at least four independent measurements. The high $D_{50}$ group showed a mean droplet diameter of 187.4 ± 21.4 µm, while the low $D_{50}$ group showed a significantly smaller mean diameter of 44.21 ± 5.6 µm. This difference was statistically confirmed using a non-parametric Mann–Whitney U test (p = 0.0159), indicating a significant difference in aerosol performance between these device subgroups. These results were presented in a bar chart **Figure 13**, with the error bars representing the standard deviations and a significance marker highlighting the difference between the groups.

This analysis highlights the critical impact of device variability, including mechanical factors such as nozzle pressure, on the aerosolization profile of the exogenous surfactant.

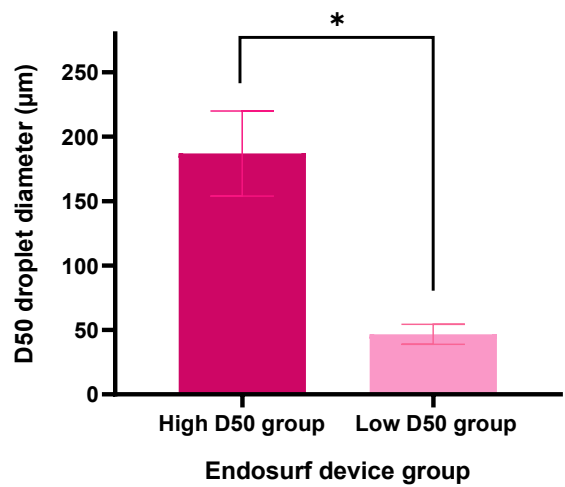

**Figure 12.** Comparison of $D_{50}$ droplet diameters between Endosurf device groups. *Bar graph comparing the median $D_{50}$ values of aerosol droplet diameters generated by Endosurf devices grouped by performance: High $D_{50}$ group (devices 4, 5, 6, 7) and Low $D_{50}$ group (devices 8, 9, 10, 11, 12). The Mann–Whitney U test showed a statistically significant reduction in droplet size in the Low $D_{50}$ group (\*p = 0.0159). Error bars represent standard deviations.*

***Ex vivo* Imaging of surfactant distribution in the lungs**

**Figure 13** shows examples of MRI images taken on isolated thoraxes before and after administration of the surfactant solution. In the images (a) taken before surfactant administration, the lung parenchyma appears hypointense due to its low tissue density compared to the chest musculature. The images in column (b) taken after surfactant administration show a strong enhancement of signal intensity in the lungs due to the presence of contrast agents in the surfactant solution. The images in column (c) correspond to a maximum intensity projection (MIP) of the 3D MRI image as seen from the front of the thorax. In these two preliminary MRI datasets, the MIP image (top panel) shows a homogeneous, distal distribution of surfactant in the lung after administration of surfactant with the Endosurf spray device. In comparison, the MIP image taken after instillation of surfactant using the LISA method (lower panel) shows a larger distribution of surfactant in the airways. These results are consistent with findings reported in recent preclinical studies exploring MRI-based visualization of pulmonary surfactant distribution in neonatal models [49].

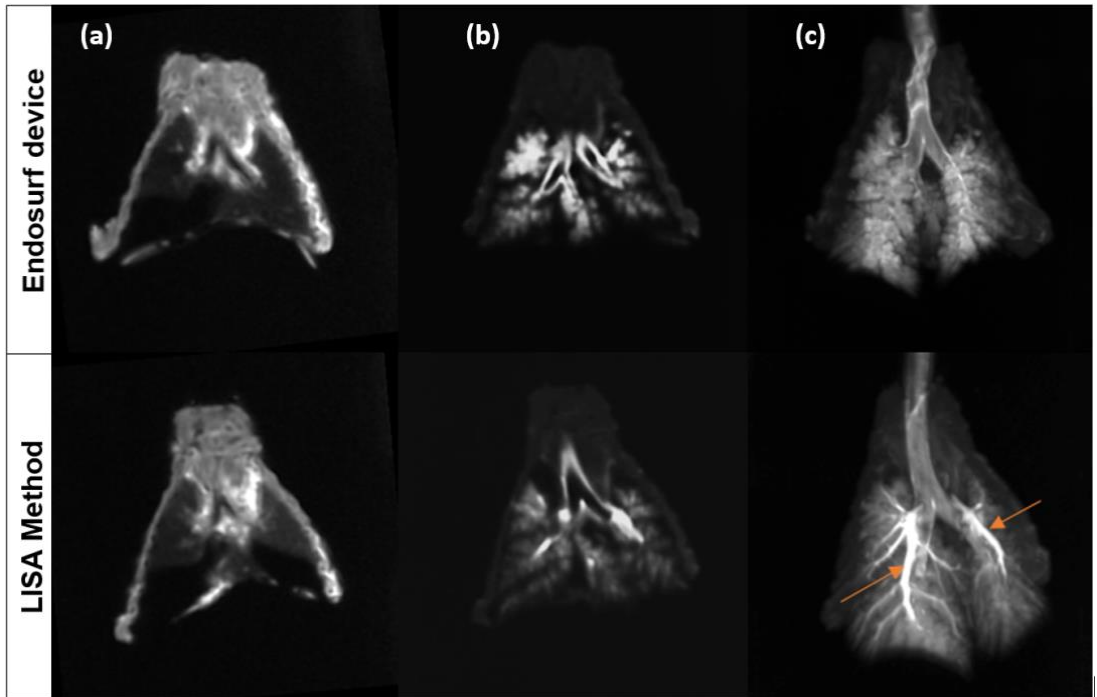

*Figure 12:* 1-mm thick coronal slices of rabbit thoraces extracted from a 3D MRI dataset acquired (a) before and (b) after administration of a solution of contrast-enhanced surfactant solution, using the Endosurf spray device (upper panel) and the LISA method (lower panel). The corresponding maximum intensity projections from the 3D data set are shown in column (c) with orange arrows pointing signal hyperintensities in the airways.

# Conclusion

This study provides important insights into the physicochemical, rheological and aerodynamic properties of Curosurf before and after atomization using different Endosurf prototypes. Atomization did not significantly alter the surface tension of the surfactant, confirming its physicochemical stability for therapeutic use. Rheological analysis showed that the viscoelastic properties of the atomized samples were comparable to those of freshly prepared Curosurf, indicating that atomization preserves the functional integrity of the surfactant. In addition, phase contrast microscopy showed a significant reduction in vesicle size after atomization, accompanied by an increase in the number of smaller vesicles, resulting in a more homogeneous size distribution that allows for better distribution in the lung. Aerodynamic studies showed that aerosol droplet size varied between the different Endosurf prototypes, with droplet diameter inversely correlated with applied pressure, highlighting the need for optimization of the device to improve aerosol generation efficiency. Preliminary *ex vivo* MRI acquisitions indicated a uniform distribution of surfactant in lung tissue following delivery *via* the Endosurf spray device, in contrast to the centralized deposition in the airways observed with the LISA method. However, further MRI acquisitions on animal lungs are required to confirm these preliminary findings and to establish a quantitative and comparative evaluation of the surfactant delivery techniques. In summary, these results confirm the potential of Endosurf devices for Curosurf surfactant delivery in the treatment of respiratory distress and support further development and clinical optimization.

**Data and materials availability**

All main data are available in the main text or in the supplementary materials. Complementary information upon data used in the analysis are available on reasonable request to corresponding author.

**Author contributions**

G.K. conducted the experiments and worked on formal analyses, wrote the original draft and contributed the review of the manuscript. J.F.B. helped for the experiments, helped for the experiments, conceptualization, data curation and contributed the review of the manuscript. Y.C. supervised the work conceptualization, data curation and contributed the review of the manuscript. N. P. supervised the work and contributed the review of the manuscript. Mu.F. methodology and contributed the review of the manuscript. Z.B. supervised the work and contributed the review of the manuscript. Mi.F. helped for the experiments, conceptualization, helped for the methodology and contributed the review of the manuscript. R.G. helped for the experiments and contributed the review of the manuscript. E.D.D.L.R. helped for the experiments, conceptualization, helped for the methodology and contributed the review of the manuscript. L.L. helped for the experiments, helped for the methodology, conceptualization and contributed the review of the manuscript. S.P. helped for the experiments, helped for the methodology, conceptualization and contributed the review of the manuscript. J.P. designed the experiments, conceptualization, supervised the work and contributed the review of the manuscript.